\documentclass[conference]{IEEEtran}
\IEEEoverridecommandlockouts
\usepackage{cite}
\usepackage{amsmath,amssymb,amsfonts}

\usepackage{bm}
\usepackage{algorithmic}
\usepackage[caption=false,font=footnotesize]{subfig}
\usepackage{graphicx}
\usepackage{textcomp}
\usepackage{xcolor}

\def\BibTeX{{\rm B\kern-.05em{\sc i\kern-.025em b}\kern-.08em
    T\kern-.1667em\lower.7ex\hbox{E}\kern-.125emX}}
\begin{document}

\title{Near-Field User Location Inference From Far-Field Power Measurements
}

\author{
\IEEEauthorblockN{Shima~Mashhadi, Tiep~M.~Hoang, and Alireza~Vahid}
\IEEEauthorblockA{
\textit{Department of Electrical \& Microelectronic Engineering, Rochester Institute of Technology, Rochester, NY, USA}\\
\{sm3709, tmheme, arveme\}@rit.edu
}
}

\maketitle

\begin{abstract}
 Near-field beamfocusing enabled by extremely large-aperture arrays (ELAA) is a promising 6G technique for massive connectivity and high spectrum efficiency. While beamfocusing concentrates energy at an intended user, the radiated field outside the focal point exhibits a structured leakage that varies with the focal-point coordinates. This paper shows that this leakage enables a new form of passive user localization in which distributed far-field sensors measuring only received power can infer the user’s location by exploiting this location-dependent power signature.
 Using the induced noncentral chi-square
statistics, we derive a Bayesian Cramér-Rao lower bound (BCRLB) that establishes the fundamental limits of this inference problem. We then evaluate a model-based grid-search estimator and an attention-based permutation-invariant deep learning regressor (DeepSet). Results under both line-of-sight (LoS) and multipath propagation confirm that reliable location inference is feasible, with accuracy improving as more sensors and snapshots are used.
\end{abstract}


\section{Introduction}
Future next-generation cellular systems (6G) are moving toward higher frequency bands (FR3, 7--24~GHz) and extremely large aperture arrays (ELAAs), which shift operation from conventional far-field (FF) propagation to the radiative near-field (Fresnel) regime with spherical wavefronts~\cite{bjornson2024towards}. In this regime, the base station (BS) can perform \emph{beamfocusing}, i.e., concentrate energy not only in angle but also at a specific focal point in space corresponding to a user equipment (UE) location. This capability is widely recognized as foundational for 6G~\cite{Nguyen20226G}. Meter-scale apertures are also essential to overcome severe path loss at high frequencies. In such settings, the Rayleigh distance can reach tens to hundreds of meters, making near-field (NF) communication central to XL-MIMO design~\cite{bjornson2024towards,bjornson2024enabling}.

Prior studies have quantified performance gains of beamfocusing~\cite{zhang2022beam, mashhadi2025near,10976082} and advanced NF systems via comprehensive channel models~\cite{Cui2022channelnf} and polar-domain codebook designs for efficient beam training~\cite{Liu2024Polar}. Beyond communications, NF beamfocusing is also pursued for integrated sensing and communications (ISAC), where the BS intentionally shapes spatially structured wavefronts to improve parameter estimation in the Fresnel regime~\cite{Sun2025ISAC,Yang2025isacfocus}. Collectively, this literature establishes beamfocusing as a mature and actively developing technology with clear pathways toward 6G implementation.

A key physical consequence of NF beamfocusing is that the radiated field \emph{outside} the focal point depends systematically on the focal range and angle. This behavior has been studied through NF degrees-of-freedom analyses~\cite{Zhang2023inter,Kosasih2025DoF}, where spherical wavefront propagation can be interpreted as a superposition of plane waves, implying that multiple spatial frequencies are excited. Consequently, the leakage of a beamfocused transmission spreads across spatial frequencies in a structured, range-dependent manner. Related effects have also been considered in mixed NF/FF systems and NF physical-layer security (PLS), mainly from the perspective of message confidentiality against eavesdroppers~\cite{Liu2026PlS,Vtc2022Pls}. However, the inference implication of this structured leakage, namely \emph{location inference} from radiated power patterns, has not been explicitly characterized. 

In this paper, we show that NF beamfocusing can unintentionally create a \emph{location-dependent FF power signature} that enables passive UE localization. Specifically, even without access to channel state information (CSI), an NF maximum-ratio transmission (MRT) beam induces a deterministic FF power pattern that is a function of the BS focal point. This opens a new inference mechanism, where an external observer can deploy $K$ distributed FF sensors that record \emph{power-only} measurements over $L$ snapshots and estimate the UE location by inferring the BS focal point. Beyond its value for localization, this same inference mechanism raises a potential privacy concern in adversarial settings, a consideration particularly relevant as beamfocusing is actively pursued for both communication-only and ISAC applications in 6G.
 
We consider an ELAA-enabled BS serving an NF UE via MRT while $K$ passive FF sensors collect power-only observations over $L$ snapshots. We adopt a power-only observation model based on established Fresnel leakage expressions that map the focal point to an FF power pattern. Using the resulting noncentral chi-square statistics, we derive a Bayesian Cram\'er--Rao lower bound (BCRLB) for range--angle inference. We then evaluate both a model-based grid-search estimator and a permutation-invariant DeepSets regressor with attention. Finally, we validate the conclusions under both line-of-sight (LoS) and multipath propagation, showing that reliable inference is feasible, while higher accuracy requires more sensors and/or more temporal snapshots.

\section{Signal Model}\label{signal-model}

We consider a BS equipped with an ELAA of $N$-element antenna deployed as a uniform linear array (ULA) that serves a single-antenna UE located in its Fresnel region.  
In addition, $K$ single-antenna sensors are deployed in the FF of the BS. Each sensor is capable of measuring only the received power of the signal emitted by the BS. These sensors do not participate in the communication directly but act as observers, passively collecting information. Let $D=(N-1)\Delta$ denote the array aperture, and define the Rayleigh distance as $d_F \triangleq 2D^2/\lambda$.

The BS ULA is placed along the $y$-axis with its center at the origin. The $n$-th antenna element is located at \(
i_n \Delta\), where \( i_n = n - \frac{N+1}{2}\) for \( n=1,\ldots,N\), and $\Delta = \frac{\lambda}{2}$ denotes the antenna spacing. The UE location is parameterized by \(\bm{\psi} = [d, \phi]\), where \(\phi\) is the angle of departure (AoD) from the array center and \(d\) is the range from the BS's array center. We consider a UE located in the Fresnel region where beamfocusing is feasible. The distance between the UE and
the $n$-th BS element is 
\begin{equation}
d_n = \sqrt{d^2 + (i_n\Delta)^2 - 2d\,i_n\Delta\cos\phi }.
\end{equation}
Accordingly, the LoS channel between the BS and UE is\footnote{Unless stated otherwise, Sections~II--IV use the LoS models to derive the leakage mapping and bounds. Section~V validates the conclusions under multipath using the \emph{Sionna} ray-tracing simulator.}
\begin{equation}
\bm{h}_{bu}(\bm{\psi})
=
\sqrt{\beta_u}\,e^{-j\frac{2\pi}{\lambda}d}\,\bm{b}(\phi,d),
\end{equation}
where $\beta_u = (\frac{\lambda}{4\pi d})^2$ is the large-scale fading coefficient where $\lambda$ is the carrier wavelength, and $\bm{b}(\phi,d)\in\mathbb{C}^{N\times1} $ is the NF steering vector, given by~\cite{Zhang2023inter}

\begin{equation}
\begin{aligned}
\relax[\bm{b}(\phi,d)]_n
&= e^{-j \frac{2\pi}{\lambda} (d_n - d)} \\ &=
e^{-j\frac{2\pi}{\lambda}\left(\frac{(i_n\Delta)^2}{2d}\sin^2\phi-i_n\Delta\cos\phi\right)}.
\end{aligned}
\end{equation}

Assuming that the BS applies NF MRT toward the UE. The BS beamforming vector is denoted by $\bm{w}(\bm{\psi})=\frac{1}{\sqrt{N}}\bm{b}(\phi,d)$, which depends on the location of the user (\(\psi = [d, \phi]\)). 

In addition to the UE, we also consider $K$ single-antenna sensors deployed in the FF of the BS. Sensor $k$ is located at known parameters $(d_k,\theta_k)$ with $d_k$ ($d_k \ge d_F$) denoting
its distance from the BS array center and $\theta_k$ is the direction of sensor $k$ as seen from the BS array center. The channel between the BS and the $k$-th sensor is modeled as FF LoS
\begin{equation}
\bm{h}_{bs,k}
=
\sqrt{\beta_k}\,e^{-j\frac{2\pi}{\lambda}d_k}\,\bm{a}(\theta_k),
\end{equation}
where $\beta_k = (\frac{\lambda}{4\pi d_k})^2$ is the large-scale fading coefficient, and the FF steering vector is
\begin{equation}
\bm{a}(\theta_k)
=
[
e^{j\frac{2\pi}{\lambda}(\frac{1-N}{2}\Delta\cos\theta_k)},\ldots,
e^{j\frac{2\pi}{\lambda}(\frac{N-1}{2}\Delta\cos\theta_k)}
]^{\top}\!\in\!\mathbb{C}^{N\times1}
\end{equation}
The baseband signal observed at the $k$th sensor at time step $t$ is given by
\[
y_{s,k}[t] =\sqrt{P_t}\mathbf{h}_{bs,k}^H \bm{w}(\bm{\psi})\, x[t] + n_k[t],
\]
where $\bm{w} (\bm{\psi})$ is the BS beamfocusing vector designed for the UE, $x[t]$ is the transmitted symbol to UE at time $t$ which satisfy \(|x[t]|^2 = 1\). The $n_k[t]\sim\mathcal{CN}(0,\sigma^2)$ is additive white Gaussian noise at the $k$th sensor. The \(P_t\) is the BS transmit power.

\subsection{Observation Model and Assumptions}\label{osubsec:bsassumption}
We adopt a block model with $L$ snapshots, where the UE location $\bm{\psi}=[d,\phi]^{\top}$, the sensor geometry $\{(d_k,\theta_k)\}_{k=1}^{K}$, and the BS beam $\bm{w}(\bm{\psi})$ and the BS transmit power $P_t$ remain fixed within the block. The BS estimates the UE near-field channel using a standard channel estimation procedure, and then applies NF MRT.

For the inference problem, we consider $K$ single-antenna FF sensors that passively observe only received signal power. Each sensor knows its own location $(d_k,\theta_k)$ and public system parameters, including the carrier frequency, BS array geometry, and an effective transmit-power scale $P_t$ (e.g., from publicly available Federal Communications Commission (FCC) licensing information~\cite{fcc_uls_license_search,  fcc_asrn_within_radius}), and estimates $\sigma^2$ during calibration periods. The goal is to infer $\bm{\psi}$ by matching the spatial pattern of power observations across sensors. The received power at the $k$th sensor can be expressed as
\begin{equation}\label{Power}
p_k[t]\triangleq |y_{s,k}[t]|^2= |\sqrt{P_t\beta_k}\bm{a}(\theta_k)^H \bm{w} (\bm{\psi})\, x[t] + n_k[t]|^2,
\end{equation}
Under the block model, the BS beam $\bm{w}(\bm{\psi})$ and the UE location $\bm{\psi}=[d,\phi]^{\top}$ are constant over $L$ snapshots. Accordingly, the expectation of the received power is
\begin{equation}
\mathbb{E}\!\left[p_k[t]|\bm{\psi}\right]
= P_t\beta_k\,\big|\bm{a}(\theta_k)^{H}\bm{w}(\bm{\psi})\big|^2+\sigma^2,
\label{eq:Ez_given_psi}
\end{equation}
where the expectation is over the receiver noise\footnote{The symbol phase is irrelevant for the power and we consider $|x[t]|^2=1$.}.
We form the sample-mean power as
$\bar{p}_k \triangleq \frac{1}{L}\sum_{t=1}^{L} p_k[t]$,
and normalize it by the estimated noise variance, the unbiased estimation of the normalized received power at the $k$th sensor is
\begin{equation}
z_k \triangleq \frac{\bar{p}_k}{\sigma^2}-1.
\label{eq:pk_def}
\end{equation}
From \eqref{eq:Ez_given_psi}, we have $\mathbb{E}[z_k|\bm{\psi}] = \frac{P_t\beta_k}{\sigma^2}\big|\bm{a}(\theta_k)^{H}\bm{w}(\bm{\psi})\big|^2$.

\subsection{Leakage model}
With NF MRT, $\bm{w}(\bm{\psi})=\frac{1}{\sqrt{N}}\bm{b}(\phi,d)$, the power leaked toward sensor $k$ is
\begin{equation}
\big|\bm{a}(\theta_k)^{H}\bm{w}(\bm{\psi})\big|^2
= \frac{1}{N}\big|\bm{a}(\theta_k)^{H}\bm{b}(\phi,d)\big|^2
\triangleq g_k(\bm{\psi}),
\label{eq:gk_def}
\end{equation}
where $g_k(\bm{\psi})$ admits a closed-form expression based on NF array response as follows\cite{Kosasih2025DoF,Zhang2023inter}, \begin{align}
g_k(\bm{\psi})
&= \frac{N}{4\beta_1^2}
\Big[\big( C(\beta_1-\beta_2)+C(\beta_1+\beta_2) \big)^2 \nonumber\\
&\quad + \big( S(\beta_1-\beta_2)+S(\beta_1+\beta_2) \big)^2\Big] .
\label{eq:g_fresnel}
\end{align}
with $\beta_1=\frac{N}{2}\sqrt{\frac{8}{d\lambda}}\,\Delta$, $\beta_2=(\cos\theta_k-\cos\phi)\sqrt{\frac{8d}{\lambda}},
\label{eq:beta12}
$
and $C(\cdot)$ and $S(\cdot)$ denoting the Fresnel cosine and sine integrals, respectively. We use \eqref{eq:g_fresnel} as a deterministic mapping from UE location $(d,\phi)$ to the leakage pattern observed across sensors.
The key point is that $\{g_k(\bm{\psi})\}_{k=1}^{K}$ produces a distinct spatial leakage signature across sensors that varies with $(d,\phi)$, enabling localization from power-only measurements.

\subsection{Problem Statement}

Let $\bm{z} = [z_1, \dots, z_K]^\top \in \mathbb{R}^K$ denote the vector of normalized power measurements from $K$ sensors, where each $z_k$ is defined in \eqref{eq:pk_def}. Our objective is to estimate the UE location parameters $\bm{\psi} = [d, \phi]^\top$ from $\bm{z}$. The noiseless leakage pattern is characterized by $\bm{g}(\bm{\psi}) = \frac{1}{\sigma^2}[P_t\beta_1 g_1(\bm{\psi}), \dots, P_t\beta_K g_K(\bm{\psi})]^\top$, with $g_k(\bm{\psi})$ from \eqref{eq:g_fresnel}. The estimation problem thus matches observations to the model via the least-squares estimator
\begin{equation}
\hat{\bm{\psi}} = \arg\min_{\bm{\psi}\in\mathcal{C}} \|\bm{z}-\bm{g}(\bm{\psi})\|_2^2,
\label{eq:ls_estimator}
\end{equation}
where $\mathcal{C}$ denotes the feasible parameter set. Section~\ref{sec:bcrlb} derives the Bayesian CRLB for this estimation.

\section{Bayesian Cram\'er-Rao Lower Bound}\label{sec:bcrlb}

In addition to the sample-mean statistic $\bm{z_k}$ in \eqref{eq:pk_def}, we consider the normalized
\emph{instantaneous} power samples collected over $L$ snapshots. Specifically, for sensor $k$ and snapshot $t$,
\begin{equation}
z_{k,t}\triangleq \frac{p_{k}[t]}{\sigma^2},\qquad k=1,\ldots,K,\; t=1,\ldots,L,
\label{eq:zkt_def}
\end{equation}
and we stack the measurements as $\bm{z}_t\triangleq[z_{1,t},\ldots,z_{K,t}]^{\top}$ and
$\mathbf{Z}\triangleq[\bm{z}_1,\ldots,\bm{z}_L]\in\mathbb{R}^{K\times L}$.
Under the block model in Section~\ref{signal-model}, $\bm{\psi}$ is fixed over the $L$ snapshots and the
samples $\{z_{k,t}\}$ are independent across $k$ and $t$ conditioned on $\bm{\psi}$. We consider $z_{k,t}$ our statistic variable which follows a noncentral chi-squared distribution with $2$ DoF ($\chi'^2_{2}\!\left(\rho_k(\bm{\psi})\right)$), and noncentrality parameter depending on $\bm{\psi}$ as follows:
\begin{equation} \rho_k\; = \frac{2}{\sigma^2}\,P_t\,\beta_k\, g_k(\bm{\psi}), \label{eq:lambda_k_def} \end{equation} 
Let $f_2(z_{k,t};\rho_k)$ denote the likelihood of the observation $z_{k,t}$ parameterized by $\rho_k$. The log-likelihood of $z_{k,t}$ given $\rho_k$ is~\cite{johnson1995continuous}
\begin{equation}\label{logl}
\log f_{2}(z_{k,t};\rho_k)
= -\log 2 - \frac{\rho_k+z_{k,t}}{2}
+ \log I_{0}\!\big(\sqrt{\rho_k z_{k,t}}\big).
\end{equation}

Then the score function for $f_{2}(z_{k,t};\rho_k)$ is
\begin{equation} l(z_{k,t};\rho_k)\;\triangleq\;-\tfrac12 \; \;+\; \tfrac12 \sqrt{\frac{z_{k,t}}{\rho_k}}\,R_2\!\left(\sqrt{\rho_k z_{k,t}}\right) .
\end{equation}
where $R_2(t) \;\triangleq\;I_{1}(t)/I_{0}(t)$, and $I_{\nu}(t)$ is the modified-Bessel function of order $\nu$. The (per-sensor) Fisher information in $\rho_k$ is~\cite[Eq.~29.39b, p.~453]{johnson1995continuous}
\begin{equation}
J_2(\rho_k)\;=\;\mathrm{Var}\!\big[l(z_{k,t};\rho_k)\big] = \frac14(\rho_k^{-1}\eta-1).  \label{eq:J_def} 
\end{equation} 
where we define $\eta=\mathbb{E}\{z_kR_2^2(\sqrt{\rho_kz_{k,t}})\} $
which must be obtained by numerical integration.

In the Fresnel operating regime, the UE range is known \emph{a priori} to lie in a finite interval,
e.g., $d\in[d_{\min},d_{\max}]$ with $d_{\max}<d_F$.
We capture this side information by modeling $\bm{\psi}=[d,\phi]^{\top}$ as a random parameter with a prior on $d$, while $\phi$ is treated as unknown.
The BCRLB then provides an
algorithm-independent lower bound on the error covariance of any estimator of $\bm{\psi}$ as follows~\cite{van2008bayesian}.
\begin{equation}
\mathbf{C}_{\mathrm{B}} \;\triangleq\; \mathbb{E}\!\left[(\hat{\bm{\psi}}-\bm{\psi})(\hat{\bm{\psi}}-\bm{\psi})^{\top}\right]
\;\succeq\; \mathbf{J}_{\mathrm{B}}(\bm{\psi})^{-1},
\end{equation}
where $\mathbf{J}_B$ is the Bayesian information matrix (BIM) given by\cite{van2008bayesian}:
\begin{equation}
\mathbf{J}_{\mathrm{B}} \;=\; \mathbb{E}_{\bm{\psi}}\!\left[\mathbf{J}_{\mathrm{F}}(\bm{\psi})\right] \;+\; \mathbf{J}_{\mathrm{prior}},
\label{eq:bcrlb_main}
\end{equation}

where $\mathbf{J}_{\mathrm{F}}(\bm{\psi})=\mathbb{E}_{\mathbf{Z}|\bm{\psi}}\left[(\nabla_{\bm{\psi}} \ln p(\mathbf{Z}; \bm{\psi}))^2\right]$ is the Fisher information from the likelihood of $\mathbf{Z}$ conditioned on $\bm{\psi}$, and
\(
\mathbf{J}_{\mathrm{prior}}= \mathbb{E}_{\bm{\psi}}\left[(\nabla_{\bm{\psi}} \ln p(\bm{\psi}))^2\right]
\)
captures the prior information.
To ensure the prior contribution $\mathbf{J}_{\text{prior}}$ is well-defined (with non-zero curvature), we employ a Beta distribution for the distance prior. Specifically, we model $d \sim \text{Beta}(\alpha, \beta)$ scaled to the interval $[d_{\min}, d_{\max}]$, with pdf~\cite{van2008bayesian}:
\begin{equation}
p(d) \propto \tilde{d}^{\alpha-1}(1-\tilde{d})^{\beta-1}, \quad \tilde{d} = \frac{d-d_{\min}}{d_{\max}-d_{\min}},
\end{equation}
with $\alpha,\beta > 1$ to ensure positive curvature. For the azimuth angle $\phi$, we assume a non-informative uniform prior over $[0, 2\pi]$. The joint prior factorizes as $p(\bm{\psi}) = p(d)p(\phi)$.
With this parameterization, only the $d$ component contributes to $\mathbf{J}_{\text{prior}}$, yielding:
\begin{equation}
\mathbf{J}_{\text{prior}} = \begin{bmatrix}
-\mathbb{E}_d\left[\frac{\partial^2}{\partial d^2} \ln p(d)\right] & 0 \\
0 & 0
\end{bmatrix}.
\label{eq:j_prior_matrix}
\end{equation}

Next, we specify the likelihood of $\mathbf{J}_{\mathrm{F}}(\bm{\psi})$. Assuming independence across $(k,t)$ and fixed $\bm{\psi}$ (and $\rho_k$) for $L$ snapshots, the score for $\bm{\psi}$ is
\begin{equation}
\nabla_{\bm{\psi}}\log p(\mathbf{Z}\mid\bm{\psi})
=\sum_{k=1}^{K}\Big(\nabla_{\bm{\psi}}\rho_k\Big)
\sum_{t=1}^{L} l\!(z_{k,t};\rho_k).
\label{eq:score_psi}
\end{equation}
Because $\rho_k(\bm{\psi})$ is constant over the $L$ snapshots and the $z_{k,t}$ samples are independent across $t$, the conditional FIM adds linearly over sensors and snapshots:
\begin{equation}
\mathbf{J}_{\mathrm{F}}(\bm{\psi})
=
\sum_{k=1}^{K} LJ_2\!(\rho_k)
\Big(\nabla_{\bm{\psi}}\rho_k\Big)\Big(\nabla_{\bm{\psi}}\rho_k\Big)^{\top}
\label{eq:Jf_sum}
\end{equation}

With 
$\nabla_{\bm{\psi}}\rho_k
=\frac{2P_t}{\sigma^2}\beta_k\,\nabla_{\bm{\psi}} g_k(\bm{\psi}),
$
and therefore
\begin{equation}
\mathbf{J}_{\mathrm{F}}(\bm{\psi})
=
\left(\frac{2P_t}{\sigma^2}\right)^{\!2}
\!L\!\sum_{k=1}^{K}\! J_2\!(\rho_k)\,
\beta_k^{2}\,
(\nabla_{\bm{\psi}} g_k(\bm{\psi}))(\nabla_{\bm{\psi}} g_k(\bm{\psi}))^{\top}.
\label{eq:Jf_final}
\end{equation}
Equivalently, letting the $k$th row of $\mathbf{G}(\bm{\psi})\in\mathbb{R}^{K\times 2}$ be
$\beta_k\big(\nabla_{\bm{\psi}} g_k(\bm{\psi})\big)^{\top}$ and
$\mathbf{W}(\bm{\psi})=\mathrm{diag}\!\big(J_2(\rho_1),\ldots,J_2(\rho_K)\big)$,
\begin{equation}
\mathbf{J}_{\mathrm{F}}(\bm{\psi})
=
\left(\frac{2P_t}{\sigma^2}\right)^{\!2} L\,
\mathbf{G}(\bm{\psi})^{\top}\mathbf{W}(\bm{\psi})\mathbf{G}(\bm{\psi}),
\label{eq:Jf_compact}
\end{equation}
where $\nabla_{\bm{\psi}} g_k(\bm{\psi})$ is obtained by differentiating the Fresnel-based leakage expression in \eqref{eq:g_fresnel}.
Finally, by substituting \eqref{eq:Jf_compact} into \eqref{eq:bcrlb_main} and by considering the fact that only the information on $d$ contribute to the BIM, we have the following BCRLB bound:
\begin{equation}
\mathrm{BCRLB}
=
\Big(\mathbb{E}_{d}[\mathbf{J}_{\mathrm{F}}(\bm{\psi})]+\mathbf{J}_{\mathrm{prior}}\Big)^{-1}.
\label{eq:BCRLB_final}
\end{equation}
\section{Permutation-Invariant Deep Learning}
\label{sec:dl}

We propose a permutation-invariant deep learning framework to estimate the UE location $\bm{\psi}=[d,\phi]^{\top}$ from distributed sensor observations. The training set contains pairs $\{(\mathcal{X}^{(i)},\bm{\psi}^{(i)})\}_{i=1}^{N_{\text{train}}}$, where $\mathcal{X}^{(i)}$ is an unordered set formed by the $K$ sensor measurements. Specifically, each set element is $\bm{x}_k=[z_k,\, d_k,\, \theta_k]^{\top}$, where $z_k$ is the normalized power in \eqref{eq:pk_def} and $(d_k,\theta_k)$ denotes the sensor location parameters. We assume an observer collects $\{\bm{x}_k\}_{k=1}^{K}$ during an acquisition phase with known UE locations (labels) and trains a model offline. Then during deployment, only $\{\bm{x}_k\}_{k=1}^{K}$ is available and the UE location is inferred without access to CSI or BS beamforming weights.

Since the sensor observations are unordered, the estimator should be invariant to sensor indexing. We therefore adopt a DeepSets architecture with attention~\cite{zaheer2017deep,lee2019set}. Let $\mathcal{X}=\{\bm{x}_1,\ldots,\bm{x}_K\}$ denote the input set. The model applies a shared encoder $\phi(\cdot)$ to each element, aggregates the features to enforce permutation invariance, and maps the aggregated representation to the location estimate through $\rho(\cdot)$:
\begin{equation}
\hat{\bm{\psi}}
=
\rho\!\left(\sum_{k=1}^{K} \alpha_k\,\bm{h}_k\right),
\qquad
\bm{h}_k=\phi(\bm{x}_k),
\label{eq:deepsets}
\end{equation}
where $\alpha_k$ are attention weights and all network parameters are trainable.

The attention weights are computed as~\cite{lee2019set}
\begin{equation}
\alpha_k
=
\frac{\exp\!\left(\bm{q}^{\top}\tanh(\bm{W}\bm{h}_k)\right)}
{\sum_{j=1}^{K}\exp\!\left(\bm{q}^{\top}\tanh(\bm{W}\bm{h}_j)\right)},
\label{eq:attention}
\end{equation}
where $\bm{W}$ is a weight matrix, $\bm{q}$ is a query vector, and both are trainable. The network is trained by minimizing the mean squared error between $\hat{\bm{\psi}}$ and $\bm{\psi}$ over the training set. This approach complements model-based estimators by learning the power-pattern-to-location mapping directly from data, and can accommodate model mismatch due to propagation.

\section{Numerical Results}
\label{sec:results}

We evaluate the performance of the proposed BCRLB, model-based MSE estimator, and DeepSets-based learning approach through extensive Monte Carlo simulations. This section describes the simulation setup, and provides comparative analysis.

\subsection{Simulation Setup}

We consider a BS equipped with an $N=100$ element ULA at $f_c=15$ GHz ($\lambda\approx20$ mm) and spacing $\Delta=\lambda/2$, yielding aperture $D\approx1$ m. The Fresnel distance is $d_F\approx100$ m, and the Bj{\"o}rnson distance is $d_B=2D=2$ m. The UE is located in the near-field region with range $d\in[d_B,d_F/8]$ (the beamfocusing region\cite{bjornson2024towards}) and azimuth $\phi\in[\pi/6,5\pi/6]$. We deploy $K$ passive single-antenna sensors in the far field ($d_k\ge d_F$) at known locations; each sensor collects $L$ power-only snapshots. For the Beta prior on distance, we set $\alpha=\beta=2$, giving prior curvature $J_d^{\text{prior}}= (\alpha+\beta-2)(\alpha+\beta-1)/(d_{\max}-d_{\min})^2 \approx0.0185$. Table~\ref{tab:sim_params} lists all system parameters. Results are averaged over $1000$ independent realizations of sensor geometries and UE locations.

\begin{table}[t]
\caption{Simulation Parameters}
\label{tab:sim_params}
\centering
\footnotesize
\begin{tabular}{|l|c|}
\hline
\textbf{Parameter} & \textbf{Value} \\
\hline
Carrier frequency $f_c$ & 15 GHz \\
Array elements $N$ & 100 \\
Antenna spacing $\Delta$ & $\lambda/2$ \\
Fresnel boundary $d_F$ & $\approx 100$ m \\
UE range $d$ & $[2, 12]$ m \\
UE azimuth $\phi$ & $[\frac{\pi}{6}, \frac{5\pi}{6}]$ \\
Sensors $K$ & $\{20,40\}$ \\
Sensor ranges $d_k$ & $[100, 150]$ m \\
Snapshots $L$ & $\{1, 50\}$ \\
Transmit power $P_t$ & 23 dBm \\
Noise power $\sigma^2$ & $-65$ to $-85$ dBm \\
\hline
\end{tabular}
\end{table}

\begin{figure}[t]
    \centering
    \includegraphics[width=0.98\columnwidth]{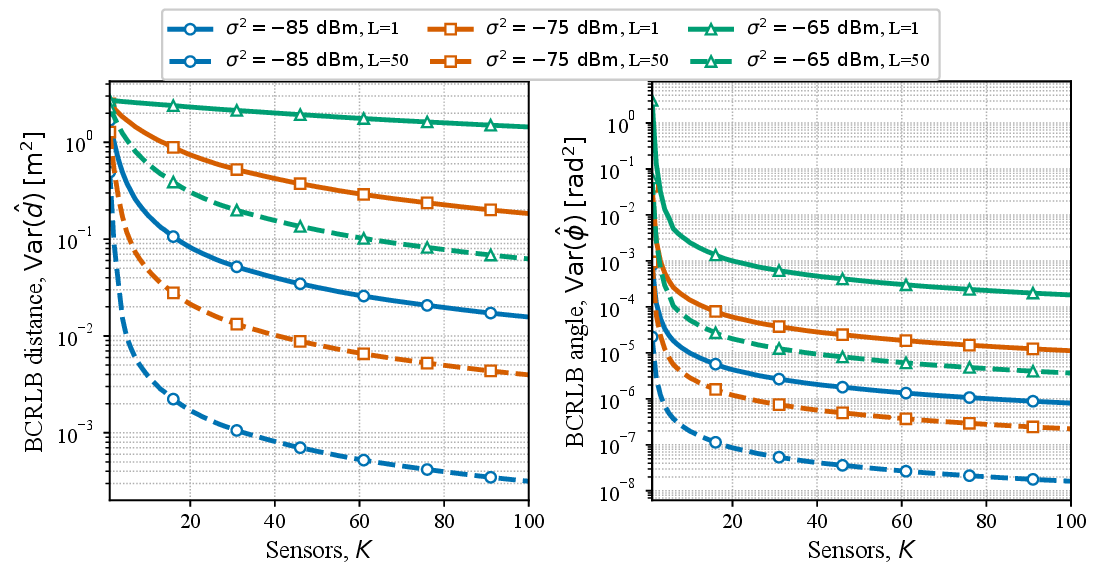}
    \label{fig:bcrlb_d}
    \caption{BCRLB versus number of passive FF sensors $K$ for different noise variances $\sigma^2$ and two snapshot lengths $L$. Solid: $L=1$; dashed: $L=50$.}
    \label{fig:bcrlb_subfigs}
\end{figure}

Fig.~\ref{fig:bcrlb_subfigs} presents the BCRLB for UE distance and angle versus the number of $K$, under different noise powers and two snapshot length.
First, angle estimation achieves lower bounds than distance across all configurations. This is because angular information is encoded in the peak direction of the leakage pattern that can be inferred even with few sensors. In contrast, range information in the NF is embedded in the spatial dispersion of the leakage across angles, a more subtle feature that requires sufficient angular sampling to resolve. 
Moreover, the prior on $d$ bounds the distance error even at high $\sigma^2$ values ($\sigma^2 = -65$ dBm corresponding to sensor $\frac{P_t\beta_k}{\sigma^2} \approx -40$ dB) and low sensors number, where measurements provide no useful information. Without this prior, the bound would be unbounded. These results confirm that the power leakage pattern inherently contains information about UE location accessible to passive FF sensors. To demonstrate the practicality of this localization mechanism, we implement two estimation approaches, grid-search MSE and DeepSets, and evaluate their performance against the BCRLB. 

\begin{figure}[t]
    \centering
    \includegraphics[width=0.6\columnwidth]{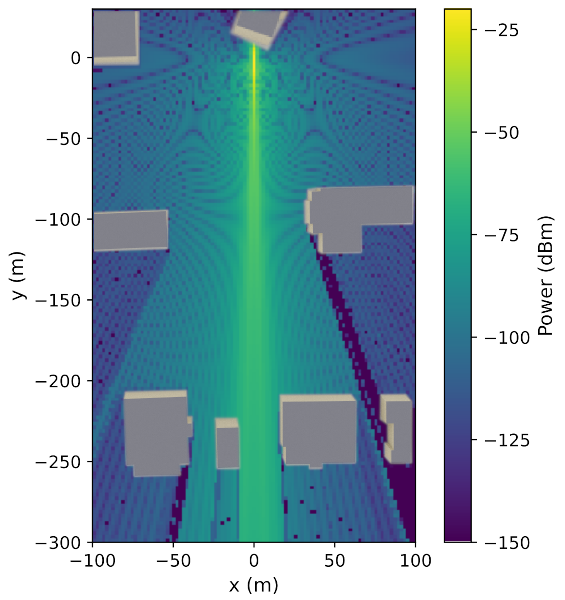}
    \caption{NF beamfocusing leakage power map (dBm) over the 2D area with building layout.}
    \label{fig:nf_beam_power}
\end{figure}
\subsection{Model-Based Grid-Search MSE Estimator}

As a model-based baseline, we implement a dense grid search over the feasible region $\mathcal{C}$. The search space is discretized into $N_d = 100$ uniformly spaced distance samples and $N_\phi = 180$ uniformly spaced angle samples. For each candidate $\bm{\psi}^{(i)}$, the predicted leakage vector $\bm{g}(\bm{\psi}^{(i)})$ is computed via \eqref{eq:g_fresnel}. The estimate is:
\begin{equation}
\hat{\bm{\psi}} = \arg\min_{\bm{\psi}^{(i)} \in \mathcal{C}_{\text{grid}}} \|\bm{z} - \bm{g}(\bm{\psi}^{(i)})\|_2^2.
\end{equation}
This approach requires no initialization or iteration, directly evaluating the MSE surface over the discretized parameter space.

\subsection{Deep Learning Setup}
\begin{table}[t]
\caption{Deep Learning Configuration}
\label{tab:dl_params}
\centering
\footnotesize
\begin{tabular}{|l|c|}
\hline
\textbf{Parameter} & \textbf{Value} \\
\hline
Architecture & DeepSets + Attention \\
Encoder $\phi(\cdot)$ & 4 layers, 256 neurons \\
Aggregation & Summation \\
Attention dimension & 256 \\
Decoder $\rho(\cdot)$ & 4 layers, [256,128,64,1] \\
Activation & GELU \\
Optimizer & Adam \\
Learning rate & $1\times10^{-3}$ \\
Batch size & 32 \\
Training samples & 35,000 \\
Validation samples & 15,000 \\
Test samples & 1,000 \\
\hline
\end{tabular}
\end{table}
The network is implemented in PyTorch and trained with Adam and early stopping based on the validation loss. We repeat the entire training process for $5$ independent \emph{sensor sets}, where each set consists of $K$ FF sensors with fixed locations. For each sensor set, we generate a dedicated dataset by sampling UE locations uniformly over $\mathcal{C}$ and computing the corresponding measurements using the model in Section~\ref{osubsec:bsassumption}, and we train a separate DeepSets model. During evaluation, we average results over $1{,}000$ test UE locations per sensor set and then average across the $5$ sensor sets. 

\begin{figure}[t]
    \centering

    \subfloat[$L=1$\label{fig:pair_L1}]{
        \includegraphics[width=0.97\linewidth]{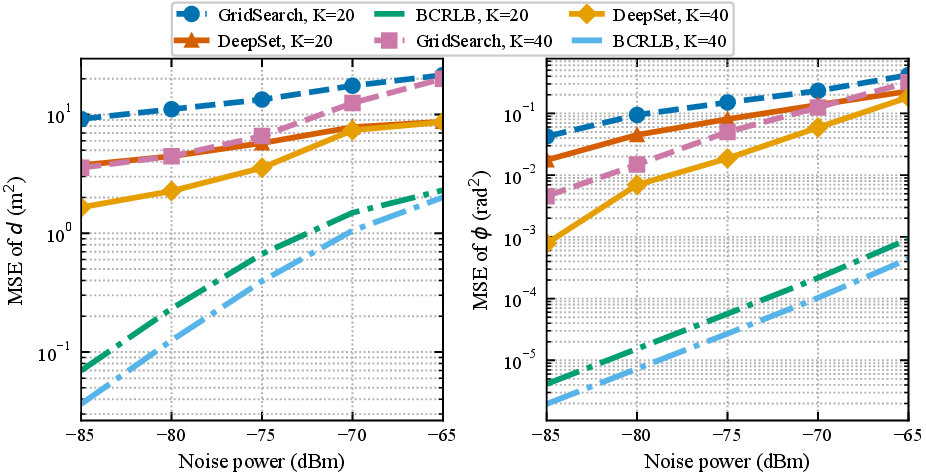}
    }\\[-0.5mm]

    \subfloat[$L=50$\label{fig:pair_L50}]{
        \includegraphics[width=0.97\linewidth]{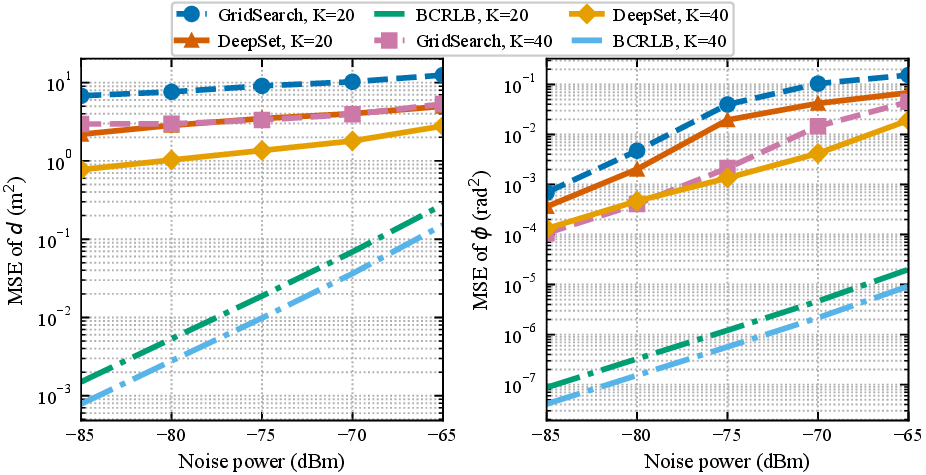}
    }
    \caption{MSE versus noise power. Here, $K$ is the number of passive FF sensors and $L$ is the number of power snapshots collected per sensor. In each subfigure, the left panel shows distance $d$ and the right panel shows angle $\phi$.}
    \label{fig:mse_pair_2x1}
\end{figure}
Fig.~\ref{fig:mse_pair_2x1} evaluates the two estimators against the BCRLB for $K \in \{20,40\}$ sensors and snapshot lengths $L=1$ and $L=50$ across multiple noise powers. Both DeepSets and grid-search achieve finite localization error that improves systematically with more sensors (denser angular sampling of the leakage field) and more snapshots, while degrading with increasing noise as expected. DeepSets consistently outperforms the dense grid-search across all configurations, achieving the best estimation errors of $0.7$ m$^2$ for range and $10^{-4}$ rad$^2$ for angle at $\sigma^2 = -85$ dBm with $K=40$ and $L=50$. At high noise with $L=1$, the distance BCRLB becomes prior-limited, indicating that the likelihood becomes uninformative. Nevertheless, the results confirm that the FF leakage pattern remains sufficiently location-dependent to enable inference of both range and angle from received power alone.

\begin{table}[t]
\caption{Sionna Ray Tracing Parameters}
\label{tab:sionna_params}
\centering
\footnotesize
\begin{tabular}{|l|c|}
\hline
\textbf{Parameter} & \textbf{Value} \\
\hline
Transmitter Antenna Pattern & TR 38.901 \\
Transmitter Polarization & Vertical \\
Receiver Antenna Pattern & Isotropic \\
Receiver Polarization & Vertical \\

 Number of the paths& $3$ \\
Line-of-Sight (LOS) & Enabled \\
\hline
Training Samples & 20,300 \\
Validation Samples & 8,700 \\
Testing Samples & 1,000 \\
\hline
\end{tabular}
\end{table}

\subsection{Performance Under Multipath Propagation}

The previous analysis assumed ideal LoS propagation, where the received power follows the closed-form leakage model \eqref{eq:g_fresnel}. To evaluate the robustness of the proposed methods under realistic propagation, we next generate data using \emph{Sionna}~\cite{hoydis2022sionna}, which captures multipath effects and therefore introduces a {model mismatch} relative to the LoS-based analytical model. For this setup, we consider a $300\times 200~\text{m}^2$ rural area near the Rochester Institute of Technology (RIT) campus in Henrietta, Rochester, NY. The BS is placed at the origin, and we distribute $K$ passive sensors in the far field of the BS. The simulated environment layout and the resulting NF leakage (power) map obtained from Sionna are shown in Fig.~2. Following the LoS setup, we use $5$ independent $K$-sensor sets (with LoS-feasible placements to the BS), train a separate model per set using Sionna-generated data, and report results averaged across the $5$ sets.
The Sionna specific configuration is summarized in Table~\ref{tab:sionna_params}, and the general simulation parameter is the same as Table~\ref{tab:sim_params}. 
\begin{figure}[t]
    \centering
    \includegraphics[width=0.99\linewidth]{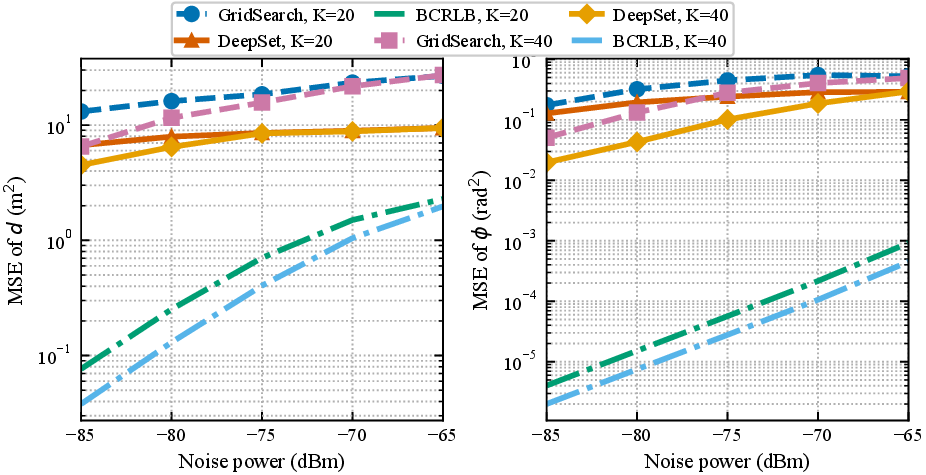} 
    \caption{Sionna-based performance under multipath propagation. MSE of the distance estimate $d$ (left) and MSE of the angle estimate $\phi$ (right) versus noise power. Here, $K$ denotes the number of passive FF sensors and $L=1$ power snapshot is used due to the computational cost of ray-tracing data generation.}
    \label{fig:sionna_results}
\end{figure}

Fig.~\ref{fig:sionna_results} shows that the Sionna-RT results follow the same overall trend as the LoS-only case, where increasing the sensing resources improves localization accuracy. However, multipath and blockages reduce the achievable gain compared to the analytical model. Because buildings restrict feasible LoS sensor locations, sensors cannot be distributed uniformly over all directions and some angular sectors contain few or no sensors, which limits the marginal benefit of adding more sensors. The grid-search uses ray-traced expected power for each candidate $\bm{\psi}$, therefore, the sensing channel is available to construct the search grid, while the learning-based method does not require explicit CSI at inference time. Due to the high computational cost of ray-tracing-based data generation, we report results only for $L=1$. Overall, the figure confirms that reliable location inference remains feasible under multipath propagation, and higher accuracy requires sufficiently many sensors and or more snapshots to mitigate noise.

\section{Conclusion}
This paper identified a new location inference mechanism in NF beamfocusing based on FF power-only measurements. We showed that a UE-directed NF beamfocusing produces a FF power pattern that depends on the focal point, enabling an external observer to estimate the UE range and azimuth from power-only measurements collected by distributed FF sensors. We provided a benchmark via a BCRLB for the resulting noncentral chi-square observation model, and demonstrated practical inference using both a grid-search estimator and an attention-based DeepSets approach. Results confirm that the inference becomes stronger as sensing resources increase (more sensors and/or more snapshots), and remains feasible under multipath propagation, which also suggests a potential privacy consideration for beamfocusing-based systems. 

\ifCLASSOPTIONcaptionsoff
\newpage
\fi

\bibliographystyle{IEEEtran}

\bibliography{references_ieee}
\end{document}